\documentclass[twocolumn,showpacs,superscriptaddress,preprintnumbers,floatfix]{revtex4}
\usepackage{latexsym}
\usepackage{amsmath}
   \def\be{\begin{equation}}
   \def\ee{\end{equation}}
   \def\ba{\begin{eqnarray}}
   \def\ea{\end{eqnarray}}

   \def\tphi{\tilde{\phi}}

   \newcommand{\fss}[1]{#1\!\!\!/}
   
   \newcommand{\qu}{\partial^2}
   
\begin{document}
\addtolength{\belowdisplayskip}{-0.3cm}       
\addtolength{\abovedisplayskip}{-0.3cm}       
\title{Gravitational corrections to Yukawa systems}
\author{O. Zanusso} 
\affiliation{
SISSA, via Beirut 4, 34014
Trieste, Italy and 
INFN , sezione di Trieste, Italy}
\author{L. Zambelli} 
\affiliation{Dipartimento di Fisica, Universit\`a degli Studi di Bologna
   and INFN sezione di Bologna, \\ via Irnerio, 46 -- I-40126 Bologna -- Italy}
\author{G. P. Vacca} 
\affiliation{Dipartimento di Fisica, Universit\`a degli Studi di Bologna
and INFN sezione di Bologna, \\ via Irnerio, 46 -- I-40126 Bologna -- Italy}
\author{R. Percacci} 
\affiliation{
SISSA, via Beirut 4, 34014
Trieste, Italy and 
INFN , sezione di Trieste, Italy}

\begin{abstract}
{\bf Abstract.}
We compute the gravitational corrections to the running of couplings in 
a scalar-fermion system, using the Wilsonian approach.
Our discussion is relevant for symmetric as well as for broken scalar phases. 
We find that the Yukawa and quartic scalar couplings become irrelevant at the
Gaussian fixed point.
\end{abstract}
\maketitle
\section{Introduction}
\vspace{-0.2cm}

The lack of renormalizability of Einstein's theory does not preclude
the possibility of calculating quantum corrections to low energy
processes due to graviton loops \cite{donoghue}.
This effective field theory approach has been applied to
calculate corrections to the gravitational potential
\cite{bdh1} and the running of Newton's constant \cite{reuter,dou,bdh2}.
Graviton loops also contribute to the beta functions of
matter couplings. This has been studied in the case of
a scalar field in \cite{griguolo}.
More recently, there has been considerable interest in
(and controversy about) the corrections to the beta function
of gauge couplings \cite{wilczek}.
Aside from the intrinsic theoretical interest, such effects
could have obvious applications to grand unified theories,
whose characteristic energy scale is not too distant from the
Planck scale, where gravity becomes strong.
In fact, it has been argued recently \cite{calmet} that
in the determination of the GUT scale, quantum gravitational effects 
could be more important than two loop effects.

With these motivations in mind, and in the same spirit,
we will calculate here the gravitational effects on the beta functions
of a simple Yukawa theory, consisting of one scalar and $N_f$ fermion fields.
We will do our calculations in flat Euclidean space, and therefore
we will not calculate here the effect that the matter has on the
running of the gravitational couplings (e.g. Newton's constant),
but at least in the limit where the matter couplings are negligible, 
this effect is easily calculable \cite{perini1}.

In addition to the above, there is also another reason for studying this problem.
If we look for a fundamental, as opposed to effective, theory of quantum gravity, 
there is now the concrete possibility that a purely field theoretic 
solution can be obtained, provided that the renormalization group has a fixed point
with a finite number of UV attractive (relevant) directions.
A theory with these properties is said to be asymptotically safe
and has the same good properties (finiteness, predictivity) as, for example, QCD.
The failure of perturbation theory means that the Gaussian fixed point
of gravity does not have the desired properties.
Work done in the last ten years has provided rather convincing evidence for the
existence of a suitable nontrivial fixed point in pure gravity;
see \cite{fp} for reviews.
It is then important to make sure that this fixed point persists
also when interacting matter is brought in.
In the case of scalar interactions, this was discussed in \cite{perini2}.
It was shown that there exists a ``Gaussian matter fixed point'',
where the gravitational couplings are nonzero and slightly shifted
relative to pure gravity, but all scalar selfinteractions are 
asymptotically free or zero.
Our results imply that such a fixed point exists also in the presence of a Yukawa coupling.

Finally we mention that asymptotic safety may play a role
also in the standard model. Some evidence for a nontrivial
fixed point in Yukawa systems has appeared recently \cite{gies}.
If this was the case, then the calculations presented here
are necessary to complete the picture by including also
the gravitational interactions.
\vspace{-0.5cm}
\section{Yukawa system}
\vspace{-0.4cm}

In this section we set up the calculation.
The flow of the renormalized couplings will be computed on a flat Euclidean background
using an exact flow equation.
An infrared cutoff, denoted $k$, is introduced via a cutoff term $\Delta S_k$,
in order to define a
scale dependent generating functional of connected Green's functions:
\be
W_k{[J]}=-\log\int [d \Phi] e^{-S[\Phi]-\int J \Phi - \Delta S_k[\Phi]}\,.
\ee   
In flat space the cutoff term has the general form
$\Delta S_k[\Phi]=\frac{1}{2}\int d^4x \,\Phi R_k^\Phi(-\partial^2)\Phi$
and $R_k^\Phi(z)$ is constructed so as to suppress the contributions
to the functional integral from the infrared modes of the field $\Phi$.
For a scalar $\phi$, we choose $R_k^\phi(z)=k^2 r(z/k^2)$,
with $r(y)=(1-y)\theta(1-y)$ \cite{Litim}, leading to the substitution
$-\partial^2 =z\to P_k(z)=z+ k^2 r(z/k^2)$, a kind of cutoff-propagator.
For a fermion $\psi$, 
$R_k^\psi(i\fss{\partial})=(\sqrt{P_k(-\partial^2)/(-\partial^2)}-1)i\fss{\partial}$.

The cutoff-corrected Legendre transform  $\Gamma_k=W_k-\int d^4x J \phi -\Delta S_k(\phi)$
defines the effective average action $\Gamma_k$ satisfying the
renormalization group equation \cite{Wetterich,Morris}
\be
\partial_t \Gamma_k=
\frac{1}{2}\,\mathrm{STr}\!\left[\left(\frac{\delta^2
   \Gamma_k}{\delta\phi \delta\phi}+R_k\right)^{-1} \partial_t R_k\right]\,,
\label{RGeq}
\ee 
where $t=\ln k$ and STr denotes a functional trace, including a factor $-1$ for fermions.
We will restrict our considerations to functionals $\Gamma_k$ of the following form
\be
\Gamma_k \!\left[g_{\mu\nu},\phi,\psi,\bar{\psi} \right] \!=\! \int\!\!
\mathrm{d}^4 x (L_b+L_f+L_g+L_{GF}+L_{gh}) 
\,.
\label{system}
\ee
The theory contains a single scalar field with Lagrangian
$$
L_b=\sqrt{g}(\tfrac{1}{2}Z_\phi\nabla^\mu\phi\nabla_\mu\phi+V(\phi))\ .
$$
We choose the potential $V$ to be even in $\phi$.
Then, there are $N_f$ Dirac fermions $\psi$ with $U(N_f)$-symmetric Lagrangian
$$
L_f=\sqrt{g}(\tfrac{i}{2}Z_\psi(\bar{\psi}\gamma^\mu D_\mu\psi-D_\mu\bar{\psi}\gamma^\mu\psi)
+i\,H(\phi)\,\bar{\psi}\,\psi)\ .
$$
The covariant derivative is
$D_\mu\psi=\partial_\mu\psi+{\frac{1}{2}}\omega_{\mu c d}J^{c d}\psi$,
and 
$D_\mu\bar\psi=\partial_\mu\bar\psi-{\frac{1}{2}}\omega_{\mu c d}\bar\psi J^{c d}$,
where $\omega_{\mu c d}$ is the spin connection and 
$J^{cd}=\frac{1}{4}[\gamma^c,\gamma^d]$ are the $O(4)$ generators.
We will choose the $O(4)$ gauge such that the vierbein is symmetric, 
so that all vierbein fluctuations can be written in terms
of the metric fluctuations and there are no $O(4)$ ghosts~\cite{Woodard}.
For the time being we keep the function $H(\phi)$ general .
On the other hand we will set
$Z_\phi=Z_\psi=1$ and neglect anomalous dimensions.

For gravity we have the Einstein-Hilbert Lagrangian 
$$
L_g= - Z \sqrt{g}R \left[g_{\mu\nu}\right]
$$
where $Z=1/(16\pi G)$.
Similarly to previous analyses, we shall work with the background-field method.
We expand around constant backgrounds, which we still denote
$g_{\mu\nu}=\delta_{\mu\nu}$, $\phi$, $\psi$ and $\bar\psi$
with corresponding fluctuations $h_{\mu\nu}$, $\varphi$, $\chi$ and $\bar\chi$.
For diffeomorphisms we fix a covariant background gauge, with gauge fixing term
$$
L_{GF}\!=\!\frac{Z}{2\alpha} \delta^{\mu\nu}F_\mu F_\nu\ ;\
F_\mu\!=\!\left(\delta_\mu^\beta \partial^\alpha
-\frac{1\!+\!\beta}{4}\delta^{\alpha\beta}\partial_\mu\right) g_{\alpha\beta}
$$
and the ghost action term consequently given by
$$
L_{gh}=\bar{c}_\mu \left(-\delta^{\mu\nu}\partial^2
+\frac{\beta-1}{2}\partial^\mu \partial^\nu
\right)c_\nu\ .
$$

We also employ the tensor decomposition
$$
h_{\mu\nu}^{\perp}+\partial_\mu v_\nu+\partial_\nu v_\mu+
\left(\partial_\mu\partial_\nu \sigma-
\tfrac{1}{4} \delta_{\mu\nu}\partial^2 \sigma\right)+\tfrac{1}{4}
\delta_{\mu\nu}h
$$
where
$\partial^\mu h_{\mu\nu}^{\perp}=\eta^{\mu\nu} h_{\mu\nu}^{\perp}=\partial^\mu
v_\mu=0$ and $h=\delta^{\mu\nu} h_{\mu\nu}$,
for tensor ($h_{\mu\nu}^\perp$), vector ($v_\mu$) and scalar ($\sigma$,$h$)
fluctuations of the metric.

The second order expansion of the Lagrangian  (\ref{system})
in the fluctuations $h^\perp_{\mu\nu}$, $v_\mu$, $\sigma$, $h$, $\varphi$,
$\chi$ and $\bar\chi$ is given by:
\begin{widetext}
\vspace{-0.3cm}
\begin{eqnarray}
{\cal L}^{(2)}=&&
-\frac{1}{4}
h_{\mu\nu}^{\perp}\left(Z \qu +V +iH\bar\psi\psi\right) h^{\perp\, \mu\nu}
-\frac{i}{16}h^T_\mu{}^\lambda\partial_\rho 
h^T_{\lambda\nu}\bar\psi\gamma^{\mu\nu\rho}\psi
\nonumber \\
&&
+\frac{1}{2}v_\mu\!\left(\!\frac{Z}{\alpha} \qu +V +iH\bar\psi\psi \!\right)\!\qu v^\mu
+\frac{i}{16}v_\mu\partial_\rho\partial^2 v_\nu\bar\psi\gamma^{\mu\nu\rho}\psi
\nonumber \\
&&
+ \frac{3}{32}\qu\sigma
\left(\frac{\alpha-3}{\alpha}Z\qu-2V-2iH\bar\psi\psi\right)\qu\sigma
+3\frac{\beta-\alpha}{16\alpha} Z\qu \sigma \qu h 
-\frac{1}{32}h\left(\frac{\beta^2-3\alpha}{\alpha}Z\qu-2V-2iH\bar\psi\psi \right)h
\nonumber \\
&&
+\frac{1}{2} (V'+iH'\bar\psi\psi) h\, \varphi+\frac{1}{2} \varphi \,(-\qu+V''+iH''\bar\psi\psi) 
\varphi
-\frac{1}{2} \bar{c}_\mu \qu c^\mu
+\frac{i}{2}\left(\bar\chi\gamma^\mu\partial_\mu\chi-\partial_\mu\bar\chi\gamma^\mu\chi\right)
+iH\bar\chi\chi 
\nonumber \\
&&
+iH'\varphi(\bar\psi\chi+\bar\chi\psi)
+\frac{i}{2}Hh(\bar\psi\chi+\bar\chi\psi)
+\left(\frac{i}{4}\partial^2v_\nu
+\frac{3i}{16}\partial_\nu\partial^2\sigma
-\frac{3i}{16}\partial_\nu h\right)
(\bar\psi\gamma^\nu\chi-\bar\chi\gamma^\nu\psi)
\nonumber
\end{eqnarray}
\vspace{0.2cm}
\end{widetext}
where the primes denote derivatives w.r.t. $\phi$
and $\gamma^{\mu\nu\rho}=\gamma^{[\mu}\gamma^\nu\gamma^{\rho]}=
\{J^{\mu\nu},\gamma^\rho\}$.
We also use the redefinitions
$-\partial^2 \sigma \to \sigma$ and $\sqrt{-\partial^2} v_\mu \to v_\mu$,
which remove the Jacobians arising from the tensor decomposition.

In order to write the RG flow for our system using (\ref{RGeq}) we
introduce a supermultiplet
$\Upsilon^T=(h_{\mu\nu}^\perp,v_\mu,c_\mu,\bar{c}_\mu,\sigma,h,\varphi,
\chi^T,\bar{\chi})$ containing all the field fluctuations of the system and
the matrix of operators
$
\Gamma_k^{(2)} = \frac{\overrightarrow{\delta}}{\delta \Upsilon^T} \Gamma_k
\frac{\overleftarrow{\delta}}{\delta \Upsilon}\,.
$
The cutoff matrix $R_k$ is chosen such that, adding it to $\Gamma_k^{(2)}$ leads to the
replacement $i\partial_\mu\to\sqrt{P_k(-\partial^2)/(-\partial^2)}i\partial_\mu$.
\section{Beta Functions.}
\vspace{-0.5cm}

Let us define the dimensionless field $\tphi=\phi/k$,
and the dimensionless functions $v(\tphi) = V(k \tphi)/k^4$ and  
$h(\tphi) =H(k \tphi)/k$.
The running of $V$ and $H$ is obtained matching
$\dot{\Gamma} \sim \int d^4x\left(\dot V+i\dot H\bar\psi\psi\right)$.
Then, $\dot v=-4v+\tphi v'+k^{-4}\dot V$, and 
$\dot h=-h+\tphi h'+k^{-1}\dot H$.
We present here the beta functionals for $v$ and $h$, in the gauge $\beta=1$
and expanding to first order in the dimensionless Newton constant $\tilde G=k^2 G$
(the full expressions are nonpolynomial in $\tilde G$):
\begin{widetext}
\vspace*{-0.3cm}
\begin{eqnarray}
\label{general}
\!\dot{v} \!\!&=&
\!\!-4 v+\tphi  v'
-\frac{N_f}{8 \pi ^2 \left(1+h^2\right)}
+\frac{3+2 v''}{32\pi^2\left(1+v''\right)}
-\tilde G\frac{ (3\!-\!\alpha) v^{\prime2}\left(2+v''\right)}{2\pi\left(1+v''\right)^2}
+\tilde G\frac{ v (3+2 \alpha)}{\pi}
+O(\tilde G^2)\,,\nonumber\\
\!\dot{h} \!\!&=& \!\!-h+\tphi h'
-\frac{h''}{32 \pi ^2 \left(1\!+\!v''\right)^2}+
\frac{h h^{\prime2}\left(2+h^2+v''\right)}{16 \pi ^2 \left(1+h^2\right)^2 \left(1\!+\!v''\right)^2}
+\tilde G\frac{(3\!-\!\alpha)v^{\prime2}}{\pi\left(1\!+\!v''\right)^3}
\left(\!\frac{1}{2} h''\left(3\!+\!v''\right)-\frac{h h^{\prime2}\left(4+3h^2+\left(2+h^2\right)v''\right)}
{\left(1+h^2\right)^2}\!\right)\nonumber\\
\!&&\!\!\!\!\!\!\!\!\!\!
+\,\tilde G h\frac{27\!+\!\alpha\left(29+96 h^2+48 h^4\right)}{16 \pi  \left(1+h^2\right)^2}
+\tilde G h' v' \frac{4\alpha\!-\!6-\!(3\!-\!2\alpha)v''+h^2(15\!-\!4\alpha)+
2 h^2 (3\!-\!\alpha)\left(\left(2\!+\!h^2\right) v''\!+\!2 h^2\right)}
{2\pi\left(1+h^2\right)^2 \left(1+v''\right)^2}
+\!O(\tilde G^2).
\end{eqnarray}
\vspace{0.2cm}
\end{widetext}
\vspace*{-0.2cm}
Fixing the form of the potentials and expanding around an appropriate basis of
operators one may find the running of any coupling of interest. 
We consider in the following local power-law potentials, 
expanding either around $\langle \tphi\rangle=0$
or $\langle \tphi \rangle=\sqrt{\kappa}$.
Concerning $h$, from now on we restrict ourselves to a simple Yukawa
interaction $h=y\tphi$.
\paragraph{\hspace{-0.2cm}Expansion around $\langle \tphi \rangle=0$. }

\hspace{-.2cm}For a quartic potential
\begin{eqnarray}
\label{vsym}
v(\tphi)&=&\lambda_0+\lambda_2\tphi^2+\lambda_4\tphi^4 \,,
\end{eqnarray}
inserting in (\ref{general}) we find, in the gauge $\alpha=0$ and in the approximation $\lambda_0=0$,
\begin{eqnarray}
\label{betasym}
\!\!\dot{\lambda}_0&=&
\frac{3+4\lambda_2}{32\pi^2\left(1+2\lambda_2\right)}-\frac{N_f}{8\pi^2}\,,
\nonumber\\
\!\!\dot{\lambda}_2&=& -2\lambda_2+\frac{N_f
 y^2}{8\pi^2}-\frac{3\lambda_4}{8\pi^2\left(1+2\lambda_2\right){}^2}+
\frac{3\tilde G\lambda_2}{\pi\left(1+2\lambda_2\right)^2}\,,
\nonumber\\
\!\!\dot{\lambda}_4&=&\frac{9\lambda_4^2}{2 \pi ^2 \left(1+2 \lambda _2\right)^3}
-\frac{N_f y ^4}{8 \pi ^2}
\nonumber\\
& &+\,3\,\tilde G\lambda_4\frac{1-10\lambda_2+36\lambda_2^2+24\lambda_2^3}{\pi
 \left(1+2 \lambda _2\right)^3}+O(\tilde G^2)\,,
\nonumber\\
\!\!\!\dot{y}&=&\frac{y^3\left(1+\lambda_2\right)}{8\pi^2\left(1+2\lambda_2\right)^2}
+\tilde Gy\frac{27+12\lambda_2\left(1+\lambda_2\right)}
{16\pi\left(1+2\lambda_2\right)^2}\,.
\label{zeroexp}
\end{eqnarray}
In general, the beta functions would depend nonpolynomially on $\lambda_0$ and $\tilde G$.
In the approximation $\lambda_0=0$, $\tilde G$ appears only polynomially: 
the highest power of $\tilde G$ occurs in $\dot\lambda_4$ and is 2. In all other terms
$\tilde G$ appears at most linearly.

When $\alpha\not=0$ one has to add the following correction terms:
\begin{eqnarray}
\label{betasymalpha}
\Delta\dot{y}&=&\alpha \tilde G y
\frac{29+180\lambda_2\left(1+\lambda_2\right)}{16\pi\left(1+2\lambda_2\right)^2}\ ,
\nonumber\\
\Delta\dot\lambda_2
&=& 2\alpha\tilde G\lambda_2\frac{1+6\lambda_2\left(1+\lambda_2\right)}
{\pi\left(1+2\lambda_2\right)^2}\ ,
\nonumber\\
\Delta\dot\lambda_4
&=&2\alpha\tilde G\lambda_4\frac{1+14\lambda_2}{\pi\left(1+2\lambda_2\right)^3}\ .
\end{eqnarray}

\paragraph{Expansion around a VEV.}
Depending on the sign of $\lambda_2$,
the potential (\ref{vsym}) can be used to describe both the symmetric and the broken phase
of the theory. In the latter case it may be more convenient to expand $v$ around
the VEV $\langle \tphi \rangle=\sqrt{\kappa}$
($\kappa \ge 0$), such that
\be
v'(\sqrt{\kappa})=0\,.
\label{derv0}
\ee
If we restrict ourselves to fourth order polynomials, $v$ has the form
\begin{eqnarray}
\label{vssb}
v(\tphi)&=&\theta_0+\theta_4(\tphi^2-\kappa)^2\ .
\end{eqnarray}
The new couplings, in the broken phase where $\lambda_2<0$, are related to those in (\ref{vsym}) by
$\theta_4=\lambda_4$,
$\kappa=-\lambda_2/2\lambda_4$,
$\theta_0=\lambda_0-\lambda_2^2/4\lambda_4$.
The beta functions of these couplings
can be derived from these relations and (\ref{betasym}).
Alternatively, one can obtain the running of $\kappa$ by deriving 
(\ref{derv0}), which yields 
\begin{equation}
\label{kdot}
\dot\kappa=-2\sqrt{\kappa}\,\dot v'(\sqrt{\kappa})/v''(\sqrt{\kappa})\ .
\end{equation}
For the broken phase, using Eq. (\ref{general}) and
retaining terms up to first order in $\tilde G$, we then obtain
\begin{eqnarray}
\label{betassb}
\!\!\dot{\theta}_0 \!\!& = \!\!&
-4\theta_0+\frac{3+16\kappa\theta_4}{32\pi^2\left(1+8\kappa\theta_4\right)}
-\frac{N_f}{8\pi^2\left(1+\kappa y^2\right)}
+\frac{3\tilde G\theta_0}{\pi}
\,,\nonumber\\
\!\!\dot{\kappa} \!\!& =\!\!&  -2\kappa+\frac{3}{16\pi^2\left(1+8\theta_4\kappa\right)^2}
-\frac{N_f y^2}{16\pi^2 (1+\kappa y^2)^2}\,,
\nonumber\\
\!\!\dot{\theta}_4 \!\!& = \!\!&
\frac{9\theta_4^2}{2\pi^2\left(1+8\kappa\theta_4\right)^3}
-\frac{N_f y^4}{8\pi^2\left(1+\kappa y^2\right)^3}
+\frac{3\tilde G\theta_4}{\pi\left(1+8\kappa\theta_4\right)^2}\,,\nonumber\\
\!\!\dot{y} \!\!& = \!\!&
\frac{y^3}{16\pi^2\left(1+\kappa y^2\right)^3\left(1+8\kappa\theta_4\right)^3}
\Bigl[2-16\kappa\theta_4\left(3+8\kappa\theta_4\right)
\nonumber\\
\!\!\!\! &&
-3\kappa y^2\left(1+8\kappa\theta_4\left(7+16\kappa\theta_4\right)\right)
-\kappa^2y^4\left(1+56\kappa\theta_4\right)
\Bigr]
\nonumber\\
\!\!\!\!  & &
\!\!\!+\frac{3 \tilde{G} y}{16 \pi  \left(1+y^2 \kappa\right)^3 \left(1+8 \theta _4 \kappa
  \right){}^2}
\Bigl[9+16 \theta _4 \kappa  \left(1+4 \theta _4 \kappa\right)\nonumber\\
&&
\!\!\!\!\!\!\! -3 y^2 \kappa  \left(1+8\theta_4 \kappa \right)\left(9+8 \theta _4 \kappa\right)
+192 y^4 \theta _4 \kappa ^3 \left(3+16 \theta _4 \kappa\right)\nonumber\\
&&\!\!\!+256 y^6 \theta _4 \kappa ^4 \left(1+4 \theta _4 \kappa \right)
\Bigr]\,.
\label{brokenvevexp}
\end{eqnarray}

We do not give here the $O(\alpha)$ corrections to these formulae. 
We notice that unlike in the expansion around $\langle \tphi \rangle =0$, 
here $\theta_0$ appears only in its own beta function.
Up to order $\tilde G$, there is no approximation involved in setting
$\theta_0=0$ in the beta functions of $\kappa$, $\theta_4$ and $y$,
as is natural in an expansion around flat space.
\section{Discussion.}

The standard $\overline{\rm MS}$ result for the beta function of the Yukawa coupling is
$\dot y=\frac{5 y^3}{16\pi^2}+\ldots$. On the other hand, neglecting
$\tilde G$ and $\lambda_2$ in (\ref{betasym}) or neglecting 
$\tilde G$ and $\kappa$ in (\ref{betassb}), we remain with
$\dot y=\frac{y^3}{8\pi^2}+\ldots$.
The difference is due to the fact that here we neglect the anomalous
dimensions of $\phi$ and $\psi$.
Since their contribution is not very small, our results are not
quantitatively accurate, but they should still give a reasonable 
qualitative picture of the gravitational corrections. 
We also stress that even though here we analyze a toy model,
our result for the leading one loop gravitational correction
applies also to realistic theories.
In particular when the Yukawa couplings form
a matrix $y_{ij}$, every beta function $\dot y_{ij}$ 
will receive the same correction $(27/16\pi)\tilde G y_{ij}$.
The inclusion of anomalous dimensions is currently under study.
Switching off the gravitational corrections, our results are in agreement
with those of \cite{gies}, when the anomalous dimensions are neglected.
Furthermore, the results for $\dot\lambda_{\,i}$ in (\ref{betasym})
are also in agreement with those of \cite{perini2}.
We have given in Eqs. (\ref{zeroexp}) and (\ref{brokenvevexp}) also the
beta functions of the vacuum energy $\lambda_0$ and $\theta_0$. 
One can see the leading contributions, proportional to $(3-4N_f)$,
the difference between the number of bosonic and fermionic degrees of freedom.

Having used an expansion around flat space, gravity is off shell.
This is the cause of the dependence of the results on
the gauge parameter $\alpha$ (and $\beta$, the dependence on which we have
computed but not reported here for simplicity).
We note that the sign of the leading corrections does not change as long as
$\alpha>0$; we have also checked that it remains the same at least
for $0\leq\beta\leq1.8$, which comprises the most popular gauge choices.
Furthermore, there are arguments showing that if $\alpha$ was allowed to run,
$\alpha=0$ would corresponds to a nonperturbative fixed point~\cite{LP}.
This suggests that the results obtained for $\alpha=0$ are probably the
most reliable.

The procedure also generically depends on the choice of 
cutoff scheme, and in particular on the cutoff function $r(y)$.
The leading terms in the beta functions of $\lambda_4$ and $y$ 
turn out to be independent on this choice, but not the
gravitational corrections, which are related to a dimensionful coupling.
In the results presented above we only used the cutoff $r(y)=(1-y)\theta(1-y)$,
so the scheme dependence is not manifest, but the numerical coefficients of the
gravitational correction would change if we used another cutoff function.
We have checked that the leading gravitational correction is
proportional to a single integral involving $r(y)$, 
so that the ratio of the leading correction terms in 
(\ref{betasym}) and (\ref{betasymalpha}) is independent of $r$.
Furthermore, the sign of the gravitational correction would be the same 
for any choice of $r(y)$ that satisfies the boundary and monotonicity
conditions to be a good cutoff.

The system (\ref{betasym}) has a (Gaussian) fixed point when $\lambda_2=\lambda_4=y=0$.
Without gravity both $\lambda_4$ and $y$ are marginal,
but the gravitational correction makes them irrelevant.
In fact the critical exponents are
$2 - (3+2\alpha)\tilde G/\pi$, 
$ -(3+2\alpha)\tilde G / \pi$ and
$-(27+29\alpha)\tilde G /16\pi $,
corresponding to the eigenvectors
$\lambda_2-3\lambda_4/16\pi^2$, $\lambda_4$ and $y$ respectively.
(Note that the gravitational corrections depend on $\alpha$ but are always negative.)
This is a remarkable result, because in the standard model
these couplings are free parameters, to be determined by experiment,
whereas here they are predicted to be zero at high energy.
Any value they have at low energy is due to the
nonlinearity of the RG flow.
This result may change in the presence of other matter fields:
it was shown in \cite{perini2} that minimally coupled matter fields
can change the sign of the critical exponent, making $\lambda_4$ relevant.
Then its value at low energy would be a free parameter,
while at high energy we would have asymptotic freedom.

All this holds both for positive and negative $\lambda_2$.
However for negative $\lambda_2$ we may obtain an improved perturbation series
\cite{gies}
expanding both $v$ and $h$ around the VEV.
Then, the beta functions are those given in (\ref{betassb}).
Most of the comments made above holds also in this case.
The main difference lies in the fact that, in the absence of
gravitational corrections, the fixed point
now has $\theta_4=y=0$ and $\kappa=3/32\pi^2$.
Remarkably, the beta function of $\kappa$ does not receive
any gravitational correction, as was already noted in~\cite{griguolo} for the
potential (\ref{vssb}) with $\theta_0=0$, even
taking into account the scalar field anomalous dimension.
This is a general property: for any scalar
potential $v$, using (\ref{kdot}) and (\ref{general}),
\vspace{0.1cm}
\ba
\, \, \, \dot{\kappa}=\!
-2\kappa+\!\frac{\sqrt{\kappa} v^{\prime\prime\prime}}{16\pi^2 v^{\prime\prime}
\left(1+v^{\prime\prime}\right)^2}
-\!\frac{h N_f\sqrt{\kappa} h^{\prime}}{2\left(1+h^2\right)^2\pi^2 
v^{\prime\prime}}\Bigg|_{\tphi=\sqrt{\kappa}}\ . \nonumber
\ea
\vspace {0.2cm}

We stress again that the beta function of $\kappa$ obtained from
the relation $\dot \kappa= -\dot{\lambda}_2/2\lambda_4+\lambda_2
\dot{\lambda}_4/2\lambda_4^2$ together with (\ref{betasym})
has a $G$ dependence in it.
Also note that the general beta functional of $h$ in (\ref{general}) 
can be used to calculate the running of any term of the form
$\phi^n\bar\psi\psi$, in particular of an explicit fermionic mass.

The gravitational corrections are of order $\tilde G=k^2/M_{\rm Planck}^2$
and therefore can be treated perturbatively at low energies. They may not be
negligible at the GUT scale, though.
Beyond the Planck scale the gravitational corrections seem to be
large and unbounded.
The theory may still be meaningful provided all couplings
(in particular $\tilde G$) reach a fixed point.
It is known that in the Einstein-Hilbert truncation gravity
has a nontrivial fixed point, also in the presence of minimally
coupled matter fields.
Since the Yukawa system has a Gaussian fixed point,
one can conclude that the theory of gravity coupled to scalars and fermions
also has a fixed point, which we may call a ``Gaussian matter'' fixed point.
However, it is clear that to study the properties of this fixed point,
in particular the critical exponents, it is necessary to calculate also the beta
function of $\tilde G$. 
There is also the possibility that the matter sector exhibits 
a nontrivial fixed point \cite{gies}.
Preliminary results indicate that, as long as $\tilde G_*\lesssim 1$, 
this fixed point would also exist in the presence of gravity.
We plan to discuss these matters in more detail elsewhere.

Acknowledgements: this work is based on the first two authors'
M.Sc. theses. O.~Z. wishes to thank A. Codello for useful advice.
R.~P. thanks D. Litim for discussions and hospitality at the Department of
Astronomy and Physics of the University of Sussex.


\end{document}